# Tunneling Magnetoresistance in Noncollinear Antiferromagnetic Tunnel Junctions

Jianting Dong[1], Xinlu Li[1], Gautam Gurung[2], Meng Zhu[1], Peina Zhang[1], Fanxing Zheng[1], Evgeny Y. Tsymbal[2], and Jia Zhang[1*]

[1]*School of Physics and Wuhan National High Magnetic Field Center, Huazhong University of Science and Technology, 430074 Wuhan, China*

[2]*Department of Physics and Astronomy & Nebraska Center for Materials and Nanoscience, University of Nebraska, Lincoln, Nebraska 68588, USA*

* jiazhang@hust.edu.cn

**Abstract**

Antiferromagnetic (AFM) spintronics has emerged as a subfield of spintronics driven by the advantages of antiferromagnets producing no stray fields and exhibiting ultrafast magnetization dynamics. The efficient method to detect an AFM order parameter, known as the Néel vector, by electric means is critical to realize concepts of AFM spintronics. Here, we demonstrate that non-collinear AFM metals, such as $Mn_3Sn$, exhibit a momentum dependent spin polarization which can be exploited in AFM tunnel junctions to detect the Néel vector. Using first-principles calculations, we predict a tunneling magnetoresistance (TMR) effect as high as 300% in AFM tunnel junctions with $Mn_3Sn$ electrodes, where the junction resistance depends on the relative orientation of their Néel vectors and exhibits four non-volatile resistance states. We argue that the spin-split band structure and the related TMR effect can also be realized in other non-collinear AFM metals like $Mn_3Ge$, $Mn_3Ga$, $Mn_3Pt$, and $Mn_3GaN$. Our work provides a robust method for detecting the Néel vector in non-collinear antiferromagnets *via* the TMR effect, which may be useful for their application in AFM spintronic devices.

Due to vanishing net magnetization, antiferromagnets produce no stray magnetic fields, exhibit high-frequency spin dynamics, and thus are promising material candidates for next-generation high-speed high-density memory devices. They play the key role in the emerging field of antiferromagnetic (AFM) spintronics [1][2], which utilizes the AFM order parameter, known as the Néel vector, as a state variable. The manipulation and detection of the Néel vector is critical for spintronic device applications. In the past few years, a significant progress has been made to manipulate the Néel vector by an electric current through the spin-torque mechanism[3][4][5][6][7]. However, due to a zero net magnetic moment in antiferromagnets, reading the AFM state out by electric means is difficult. So far, the electrical detection of the Néel vector has been performed using the anisotropic[3][4] or spin-Hall[8][9][10] magnetoresistance effects. Unfortunately, both methods suffer from relatively small signals easily influenced by perturbations[11] and require multiple in-plane terminals resulting in large device dimensions. It would be desirable to exploit the tunneling magnetoresistance (TMR) effect[12] well known for conventional CoFe/$Al_2O_3$/CoFe [13][14] and CoFeB/MgO/CoFeB [15][16][17] magnetic tunnel junctions (MTJs). Unfortunately, most AFM metals, such as L10-Mn$X$ ($X$ = Pt, Pd, and Ir), CuMnAs, $Mn_2Au$, NiO, and many others, suffer from spin degeneracy of their electronic band structures which makes TMR in AFM tunnel junctions (AFMTJs) based on these antiferromagnets unfeasible. While sizable magnetoresistive effects have been theoretically predicted for AFM spin valves[18][19] and AFMTJs[20][21], all of them relied on perfect interfaces with switchable interfacial magnetic moments, rather than on bulk properties of the antiferromagnets. This mechanism is not robust against disorder and interface roughness inevitable in experimental conditions.

     Recently, it has been predicted that there are antiferromagnets of certain magnetic space groups (MSGs) exhibiting a momentum-dependent spin splitting, even when spin-orbit coupling is absent [22][23]. Such antiferromagnets are capable of maintaining spin-polarized currents along certain crystallographic orientations [24][25][26][27] and can serve as functional electrodes in AFMTJs[28]. Among them are non-collinear antiferromagnets $Mn_3X$ ($X$ = Sn, Ge,

Ga) which belong to the D0$_{19}$ hexagonal structural phase. These materials are appealing for AFM spintronics, due to their spin-dependent transport properties, such as the anomalous Hall effect [29][30][31], the spin Hall effect [32][33], and the magnetic spin Hall effect [34], as well as the ability to generate spin polarized currents [35]. The momentum-dependent spin splitting in Mn$_3X$ antiferromagnets indicate that they can serve as electrodes in AFMTJs to produce a sizable TMR effect. On the other hand, the electric current induced switching of the non-collinear AFM order has been successfully demonstrated in Mn$_3$Sn/heavy metal bilayers at room temperature through the spin-orbit torque (SOT) mechanism [36][37]. Thus, it may be feasible to create an AFMTJ based on AFM Mn$_3X$ electrodes where the control of the Néel vector is carried out using the SOT induced switching, while its detection is performed *via* the TMR effect.

In this letter, using density functional theory and quantum conductance calculations, we predict a giant TMR effect in AFMTJs based on non-collinear Mn$_3$Sn electrodes. The effect is driven by the spin-split Fermi surface of Mn$_3$Sn producing a spin-polarized current controlled by the relative orientation of the Néel vectors in the two AFM electrodes. We argue that the momentum dependent spin splitting and the related TMR effect can also be realized in other non-collinear AFM metals providing a robust method to detect the Néel vector in these antiferromagnets *via* the TMR effect.

Fig. 1 (a) shows the atomic structure of bulk Mn$_3$Sn which belongs to the hexagonal D0$_{19}$ space group of the *P*6$_3$/*mmc* symmetry in the paramagnetic phase. Below the Néel temperature $T_N$ of 420 K, Mn$_3$Sn acquires a non-collinear AFM order where, within the *a-b* plane, Mn atoms form a Kagome-type lattice with neighboring Mn moments aligned under 120° angles with respect to each other [38]. Such a noncollinear AFM phase of bulk Mn$_3$Sn belongs to the *Cmc'm'* MSG. This MSG is characterized by space inversion symmetry (*P*), mirror reflection in the $M_b$ plane (shown in Fig. 1 (a) by the dashed line), mirror reflection in the $M_{b\perp}$ plane combined with time reversal (*T*) and a half-lattice translation along the *z*-axis ($\tau = c/2$), *i.e.* nonsymmorphic symmetry $\{TM_{b\perp}| \tau = c/2\}$, and mirror reflection in the *z*-plane ($M_z$) combined with time reversal, *i.e. TM$_z$*.

This MSG supports momentum dependent spin splitting of the electronic bands and thus a non-spin-degenerate Fermi surface. It is known that the appearance of spin splitting in antiferromagnets, not associated with spin-orbit coupling, requires the violation of both $TP\tau$ and $U\tau$ symmetries, where $U$ is spinor symmetry [23]. Due to the present space inversion but broken time reversal in AFM Mn$_3$Sn, the $TP\tau$ violation is satisfied. On the other hand, the primitive unit cell of AFM Mn$_3$Sn is equivalent to its paramagnetic unit cell. Therefore, the reversal of Mn magnetic moments by $U$ followed by translation $\tau$ cannot recover the atomic positions and the AFM order simultaneously, resulting in the $U\tau$ symmetry violation. The broken $TP\tau$ and $U\tau$ symmetries ensure that the band structure of Mn$_3$Sn is spin split. As a result, the electronic bands of Mn$_3$Sn are no longer spin degenerate and thus are expected to carry the momentum dependent spin polarization.

The noncollinear AFM order in bulk Mn$_3$Sn can be characterized by the Néel vector orientation given by angle $\alpha$, as shown in Fig. 1. There are four Néel vector orientations, $\alpha = 0°$, $60°$, $120°$, and $180°$, representing symmetry equivalent AFM states. Since the band structure of Mn$_3$Sn is spin split, the rotation of the Néel vector by angle $\alpha$ changes the spin expectation value $<s>$ at each $k$ point. The spin expectation value for each Bloch state can be calculated as follows:

$$\langle s \rangle = \frac{\hbar}{2} \langle \sigma \rangle = \frac{\hbar}{2} \langle \psi_n(k) | \sigma | \psi_n(k) \rangle \qquad (1)$$

where $\sigma$ in the Pauli matrix and $\psi_n(k)$ is the Bloch wave function.

The first-principles calculations are performed by using a Quantum ESPRESSO package by considering noncollinear magnetism but neglecting spin-orbit coupling as described in Supplementary Material[39]. We calculate the band structures and the spin expectation values for each band for the four non-collinear AFM states in Mn$_3$Sn. It is seen from Figs. 1(e-f) that for all bands, along the high-symmetry Brillouin zone directions, the spin expectation value $<\sigma>$ (indicated by color in Fig. 1 (e-f)) gradually changes with $\alpha$ changing from $0°$ to $180°$. Notably, the in-plane spin components, $<\sigma_x>$ and $<\sigma_y>$, have opposite signs for $\alpha = 0°$ and $\alpha = 180°$, indicating the reversal of the Néel vector equivalent to the time reversal symmetry operation.

This is reminiscent to ferromagnets where the magnetization reversal flips the spin of all electronic bands.

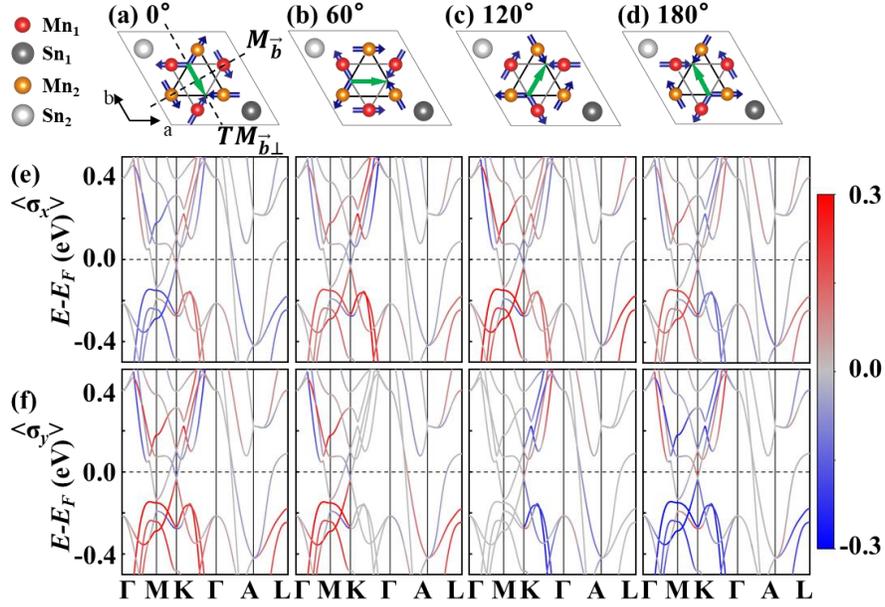

FIG. 1. (a-d) The top view of atomic and magnetic structures of AFM Mn$_3$Sn for Néel vectors (green arrows) oriented at $\alpha = 0°$ (a), $60°$ (b), $120°$ (c), and $180°$ (d). (Mn$_1$, Sn$_1$) and (Mn$_2$, Sn$_2$) layers are located at $z=c/4$ and $z=3c/4$, respectively. (e-f) The corresponding band structures along high symmetry lines in the Brillouin zone indicating in color the in-plane spin expectation values $<\sigma_x>$ (e) and $<\sigma_y>$ (f). The color scale for the spin values is shown on the right.

The shape of the Fermi surface plays a decisive role in electronic transport. Due to the spin dependent electronic band structure of Mn$_3$Sn, its Fermi surface is spin polarized. As seen from Supplementary Fig. S1, the three Fermi surface sheets, corresponding to three different bands, exhibit a complex distribution of the spin expectation values as a function of the in-plane wave vector. At the same time, the spin texture of the Fermi surface changes with the orientation of the Néel vector determined by angle $\alpha$. Fig. 2 shows the calculated spin expectation values, $<\sigma_x>$ and $<\sigma_y>$, for one of the Fermi surfaces sheets (band 1 in Fig. S1) projected to the $k_x$-$k_y$ plane for different angles α. It is seen that when the angle α changes, the projected spin texture rotates with the angle and changes its symmetry. In particular, when α is rotated from 0° to 180°, corresponding to the Néel vector reversal, the spin contrast is flipped consistent with the time reversal transformation.

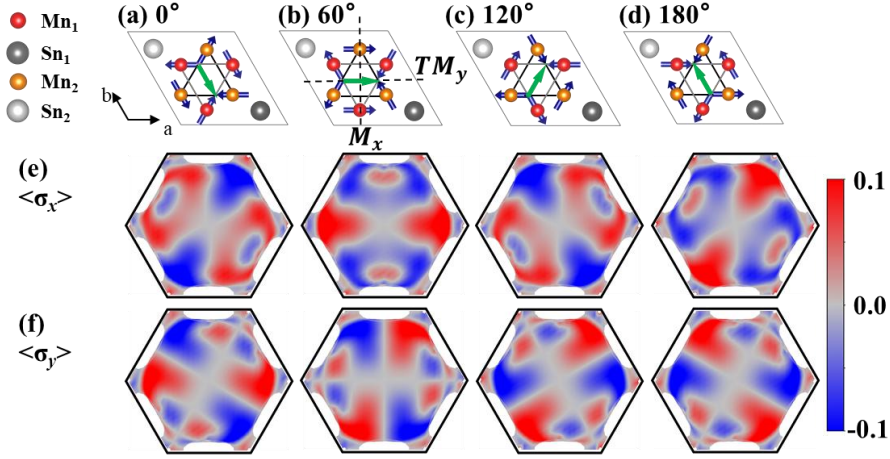

FIG. 2. (a-d) The top view of atomic and antiferromagnetic structures of bulk $Mn_3Sn$ for α = 0°, 60°, 120° and 180°. The black dash lines in (b) indicate the mirror planes perpendicular to *ab* plane for α = 60°. (e) and (f) are the in-plane spin $<\sigma_x>$ and $<\sigma_y>$ projected on the Fermi surface (band 1 in Fig. S1) of bulk $Mn_3Sn$.

The spin texture symmetry in Figs. 2(e) and (f) can be understood from the MGS of AFM $Mn_3Sn$. For instance, in the case of α = 60°, there are two mirror planes ($M_x$ and $M_y$) which are perpendicular to the *a-b* plane as indicated in Fig. 2 (b). The symmetry operations corresponding to these planes are $M_x$ and $\{TM_y | \tau = c/2\}$. In addition, there is inversion symmetry *P* and mirror reflection combined with time reversal, $TM_z$. These symmetries transform the wave vector and the spin as follows:

$P$: $(k_x, k_y, k_z) \rightarrow (-k_x, -k_y, -k_z)$; $(\sigma_x, \sigma_y, \sigma_z) \rightarrow (\sigma_x, \sigma_y, \sigma_z)$;

$M_x$: $(k_x, k_y, k_z) \rightarrow (-k_x, k_y, k_z)$; $(\sigma_x, \sigma_y, \sigma_z) \rightarrow (\sigma_x, -\sigma_y, -\sigma_z)$;

$\{TM_y|\tau = c/2\}$: $(k_x, k_y, k_z) \rightarrow (-k_x, k_y, -k_z)$; $(\sigma_x, \sigma_y, \sigma_z) \rightarrow (\sigma_x, -\sigma_y, \sigma_z)$;

$TM_z$: $(k_x, k_y, k_z) \rightarrow (-k_x, -k_y, k_z)$;   $(\sigma_x, \sigma_y, \sigma_z) \rightarrow (\sigma_x, \sigma_y, -\sigma_z)$.

By combining these symmetry restrictions, we conclude that $\sigma_x$ should be symmetric (have the same sign) with respect to $k_x$ and $k_y$, while $\sigma_y$ should be antisymmetric (have an opposite sign) with respect to $k_x$ and $k_y$. As a result, $\sigma_y$ turns out to be zero along the $k_x = 0$ and $k_y = 0$ lines. All these conclusions are consistent with the spin texture shown in Figs. 2 (e) and (f) for the case of $\alpha = 60°$.

Due to the spin-polarized Fermi surface of Mn$_3$Sn, this noncollinear AFM metal can serve as electrodes in an AFMTJ which resistance is expected to depend on the relative orientation of the Néel vectors. In particular, if the current flows along the [0001] direction (*c*-axis), *i.e.* perpendicular to the magnetic moment lying in the *a-b* plane, each propagating Bloch state (conduction channel) can be characterized by the transverse wave vector $\boldsymbol{k}_\| = (k_x, k_y)$ and the spin being oriented in the plane perpendicular to the current direction. In a fully crystalline AFMTJ with no momentum and spin-flip scattering, the transverse wave vector and thus the spin are conserved in the tunneling process. Due to the spin state being $\boldsymbol{k}_\|$ dependent and controlled by the Néel vector of the AFM metal, the electron transmission between the two AFM electrodes across the tunnel barrier should be dependent on the relative Néel vector alignment of the electrodes, as a result of spin matching (mismatching) at each $\boldsymbol{k}_\|$-point. This is similar to the conventional MTJs based of ferromagnetic electrodes, where TMR originates from electron transmission being dependent on their relative magnetization orientation.

To explicitly demonstrate the effect and estimate its magnitude, we consider an AFMTJ based on noncollinear AFM Mn$_3$Sn electrodes and a vacuum barrier layer (6 Å thick), as shown in Fig. 3(a). Due to being electronically featureless, vacuum can serve as a model tunneling barrier to explore transport phenomena focusing entirely on the spin-dependent properties of magnetic electrodes [40][41]. We calculate transmission and TMR of the Mn$_3$Sn/Vacuum/Mn$_3$Sn AFMTJ as described in Supplemental Material [39]. In the calculations, the Néel vector of the left Mn$_3$Sn electrode is kept fixed at $\alpha_L = 0°$, whereas the Néel vector on the right Mn$_3$Sn electrode is assumed to have four different orientations, $\alpha_R = 0°, 60°, 120°,$ and $180°$.

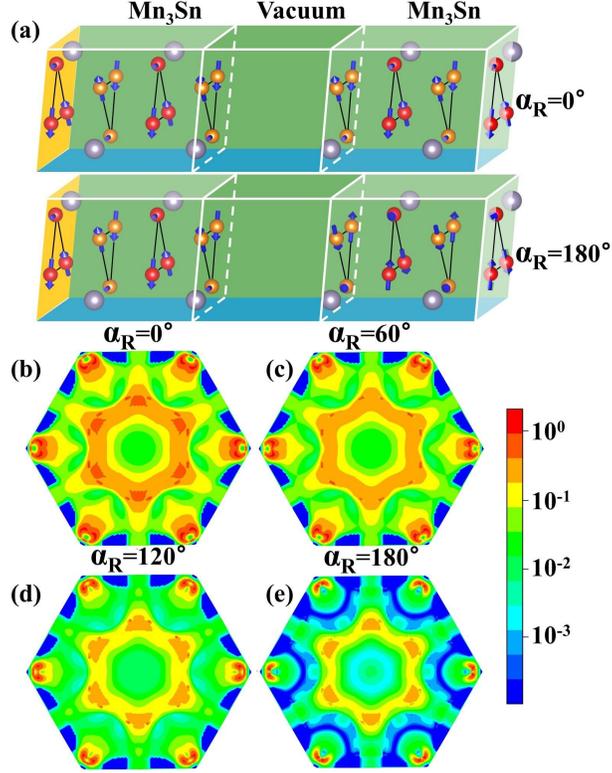

FIG. 3. (a) Schematic of the Mn$_3$Sn/Vacuum/Mn$_3$Sn AFMTJ for the parallel ($\alpha_L = 0°$, $\alpha_R = 0°$) and antiparallel ($\alpha_L = 0°$, $\alpha_R = 180°$) Néel vectors. (b-e) $k_\parallel$-resolved transmission $T(k_\parallel)$ in the 2D Brillouin zone for $\alpha_L = 0°$ and $\alpha_R = 0°$ (b), 60° (c), 120° (d), and 180° (e).

Figures 3(b-e) show the calculated $k_\parallel$ dependent electron transmission $T(k_\parallel)$ in the 2D Brillouin zone for Mn$_3$Sn/Vacuum/Mn$_3$Sn AFMTJ for a fixed $\alpha_L = 0°$ and different $\alpha_R$. The overall transmission distribution reflects the symmetry of the Mn$_3$Sn Fermi surface (Fig. S2). It is seen that for $\alpha_R = 0°$ the transmission is highest, and it decreases with increasing $\alpha_R$ from 0° to 180°. This is due the increase of $\alpha_R$ augmenting the spin mismatch between the incident electron wave from the left Mn$_3$Sn electrode and the outcoming electron wave in the right Mn$_3$Sn electrode. For $\alpha_R = 0°$, the spin state is identical in the left and right Mn$_3$Sn electrodes at each $k_\parallel$ point, whereas for $\alpha_R = 180°$, the spin state is opposite (see Fig. 2). This is similar to an MTJ with ferromagnetic electrodes where the largest transmission difference occurs between parallel ($\alpha_R = 0°$) and antiparallel ($\alpha_R = 180°$) magnetization orientations.

Fig. 4 (a) shows the tunneling conductance $G$ and resistance-area product ($RA$) of the Mn$_3$Sn/Vacuum/Mn$_3$Sn AFMTJ for a fixed $α_L = 0°$ and different values of $α_R = 0°$, 60°, 120°, and 180°. It is seen that there are four different resistance states corresponding to four different relative orientations of the Néel vector in the electrodes. Importantly, all these magnetic configurations are energetically stable and thus can be employed as non-volatile states in a spintronic device. The predicted TMR effect is comparable to that known for Fe/MgO/Fe MTJs [16][17]. Specifically, the conductance ratio between parallel and antiparallel orientations of the Néel vector is $G(α_R=0°)/G(α_R=180°) ≈ 3.9$, corresponding to the conventional TMR ratio[12] of about 300%.

We note that the predicted TMR effect is not restricted to Mn$_3$Sn but expected to occur in AFMTJs based on other non-collinear AFM metals exhibiting momentum dependent spin splitting of the Fermi surface. Among such antiferromagnets are D0$_{19}$-Mn$_3$Ge (MSG: *Cm'cm'*, $T_N$ = 380 K)[50], Mn$_3$Ga (MSG: *P6$_3$'/m'm'c*, $T_N$ = 460 K)[51], Mn$_3$Pt (MSG: *R-3m'*, $T_N$=365 K)[52] and Mn$_3$Ir (MSG: *R-3m'*, $T_N$ = 960 K)[53] in cubic phase, and antiperovskite Mn$_3$GaN (MSG: *R-3m*, $T_N$ = 298 K)[54]. The calculated spin-split band structures of Mn$_3$Ge, Mn$_3$Pt, and Mn$_3$GaN are shown in Supplemental Figs. S4-S6. The magnitude of TMR is expected to depend not only on the AFM electrodes but also on the choice of the insulating barrier layer and can be higher or lower than the predicted value for the vacuum barrier. As a demonstration, our transport calculations for a Mn$_3$Sn/1T-HfO$_2$/Mn$_3$Sn AFMTJ with a monolayer of hexagonal hafnia (1T-HfO$_2$) as a barrier predict a similarly large TMR of about 124% (Note 4 in Supplemental Material).

The predicted TMR effect can be realized in practice using an AFM spintronic device shown in Fig. 4(b), providing a fully electrical method for writing and reading out the state of the Néel vector in the non-collinear AFMTJ. Here the Néel vector of the top AFM electrode is fixed, while the Néel vector of the bottom AFM electrode can be switched by an in-plane electric current in an adjacent heavy metal through the SOT mechanism. Such a switching of the Mn$_3$Sn AFM state by the current driven SOT has been between successfully demonstrated in the recent experiments [36][37]. The state of the Néel vector in the bottom AFM electrode can be detected by measuring the tunneling resistance of the AFMTJ. The proposed three-terminal geometry is similar to that of CoFeB/MgO/CoFeB MTJs which have been extensively studied for SOT magnetic random-access memories (MRAMs) [55][56].

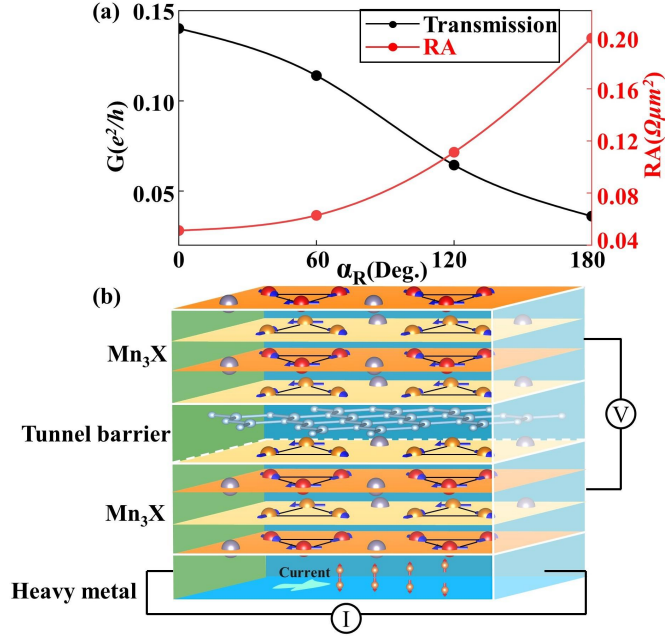

FIG. 4. (a) The calculated tunneling conductance $G$ per lateral unit cell area (left axis) and resistance-area ($RA$) product (right axis) for a $Mn_3Sn$/Vacuum/$Mn_3Sn$ AFMTJ as a function of $a_R$ = 0°, 60°, 120°, and 180°. The black and red lines are guides to the eyes. (b) Schematic of the three-terminal AFM spintronic device, where the Néel vector of the bottom non-collinear AFM electrode $Mn_3X$ ($X$ = Sn, Ge, Ga) can be switched by the electric current driven SOT and read out through the TMR effect in a $Mn_3X$/Tunnel barrier/$Mn_3X$ AFMTJ.

In summary, by performing first-principles calculations, we have investigated the features of spin-split band structures for a typical non-collinear AFM metal $Mn_3Sn$ and predicted a giant TMR effect of about 300% in a $Mn_3Sn$-based AFMTJ using a vacuum barrier and over 100% TMR using a $HfO_2$ single-layer barrier. It is possible to further enhance the TMR in the proposed AFMTJs, for instance, by engineering a tunnel barrier and non-collinear AFM electrodes. Our results provide an effective method for the electrical detection of the state of the Néel vector in noncollinear AFM metals useful for applications in AFM spintronics. The predicted TMR effect and multiple non-volatile resistance could be observe in other non-collinear AFM metals exhibiting a momentum dependent splitting of the Fermi surface. We hope that our results will stimulate experimentalists to realize the proposed AFMTJs in practice.

This work was supported by the National Natural Science Foundation of China (grant No. 12174129). The research at the University of Nebraska-Lincoln was supported by the Office of Naval Research (ONR grant N00014-20-1-2844).